\documentstyle[sprocl,epsfig,here]{article}

\def\Journal#1#2#3#4{{#1} {\bf #2}, #3 (#4)}

\def\NIM{\em Nucl. Instrum. Methods}

\def\NPB{{\em Nucl. Phys.} B}

\def\PRL{\em Phys. Rev. Lett.}

\def\ZPC{{\em Z. Phys.} C}

\newcommand{\etal}{{\it et~al.}}
\newcommand\GeV{\ifmmode {\mathrm{\ Ge\kern -0.1em V}}\else
                   \textrm{Ge\kern -0.1em V}\fi}%

\begin{document}

\title{EXPERIMENTAL CONSTRAINTS ON THE CABIBBO-KOBAYASHI-MASKAWA
MATRIX\footnote{To appear in the Proceedings of the 
{\it Workshop on CP Violation}, Adelaide, 
July 3-8 1998. }}

\author{S. MELE}

\address{CERN, CH1211, Gen\`eve 23, Switzerland\\E-mail: Salvatore.Mele@cern.ch} 


\maketitle\abstracts{The LEP investigation of the $B^0_d$ and $B^0_s$ 
oscillations and  of the Cabibbo-Kobayashi-Maskawa 
matrix element $|\mathrm{V_{ub}}|$ improve the constraints on the other
elements of this matrix.
From a fit to the experimental data and the theory calculations it is 
possible to determine the vertex of the unitarity triangle as:
\begin{displaymath}
    \rho =0.155 _{-0.105} ^{+0.115}\,\,\,\,\,
    \eta =0.383 _{-0.060} ^{+0.063}.
\end{displaymath}
The corresponding values of its angles,  in their customary definition in terms
of sines for $\alpha$ and $\beta$, are:
\begin{displaymath}
    \sin{2\alpha}  =0.08 _{-0.50}  ^{+0.43}\,\,\, 
    \sin{2\beta}   =0.75 \pm 0.10\,\,\,
    \gamma         =68 \pm 15^\circ. 
\end{displaymath}
The fit also yields indirect information on the compatibility with zero
of the CP violating phase of the matrix, 
on some non-perturbative QCD parameters and 
on the $B^0_s$ oscillation frequency.}

%
%

\vspace{-1cm}
\section{Introduction}

The  Standard Model of the electroweak interactions~\cite{sm} predicts
a mixing of the quark mass and weak interaction eigenstates, as described
 by the Cabibbo-Kobayashi-Maskawa\cite{ckm} (CKM) matrix.
This $3\times3$ unitary
matrix can be written~\cite{wolfenstein} in terms of only four real 
parameters:
\begin{equation}
  \mathrm{
    \pmatrix{ \mathrm{V_{ud}} &  \mathrm{V_{us}} &  \mathrm{V_{ub}} \cr 
      \mathrm{V_{cd}} &  \mathrm{V_{cs}} &  \mathrm{V_{cb}} \cr 
      \mathrm{V_{td}} &  \mathrm{V_{ts}} &  \mathrm{V_{tb}} \cr}
    }
  \simeq \pmatrix
  {
    1-{\lambda^2 \over 2} & \lambda &  A\lambda^3 (\rho - i\eta) \cr
    -\lambda & 1-{\lambda^2 \over 2} &  A\lambda^2 \cr
    A\lambda^3 (1-\rho - i\eta) &  -A\lambda^2 &      1 \cr
  }.
\label{equation:ckm}
\end{equation}   
$A$, $\rho$ and $\eta$ are of the order of the unity and
$\lambda$ is chosen as the sine of the Cabibbo angle. This
parametrisation, that holds to order $\lambda^4$, shows immediately the hierarchy of the couplings
of the quarks in the charged current part of the  Standard Model Lagrangian.
Moreover, the parameter
$\eta$ gives the amount of the complex phase of the matrix and is 
thus directly related to the known violation of the CP symmetry
produced by the weak interaction. The study of its compatibility with zero is
thus of great interest.

The parameters $A$ and $\lambda$ are known with an accuracy of a few 
percent and the determination
of $\rho$ and $\eta$ is presented in what follows. 
This can be achieved by means of a fit of the theory
modelling of some physical processes to the experimental data.

\begin{flushleft}
  \begin{figure}[H]
    \begin{minipage}[H]{5cm}
      \begin{center}
        \mbox{\epsfysize=1.75cm\epsffile{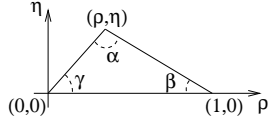}}
        \caption{The unitarity triangle.}
      \end{center}
    \end{minipage}
  \end{figure}
\end{flushleft}

\vspace{-3.0cm}
\begin{flushright}
  \begin{minipage}[H]{7cm}
    The measurement of the $\rho$ and $\eta$ parameters
    is usually associated to the determination of the only unknown vertex 
    of a triangle in the $\rho-\eta$ plane whose other two vertices are in (0,0) and (1,0). 
    This triangle,  called the unitarity triangle, is shown in Figure~1.
  \end{minipage}
\end{flushright}\vspace{-1em}

%
%

\section{Constraints}

The value of the sine of the Cabibbo angle  is known with a good accuracy~\cite{pdg} as: 
\begin{displaymath}
\lambda = 0.2196 \pm 0.0023.
\end{displaymath}
The  parameter $A$ depends on $\lambda$ and on the CKM matrix element $|\mathrm{V_{cb}}|$
and can be extracted as:
\begin{displaymath}
A = {|\mathrm{V_{cb}}|^2 \over \lambda^2} = 0.819 \pm 0.035,
\end{displaymath}
where it has been used~\cite{pdg}:
\begin{displaymath}
 |\mathrm{V_{cb}}| = (39.5 \pm 1.7)\times 10^{-3}.
\end{displaymath}

The four physical processes that show the largest sensitivity to the values of 
the CKM parameters $\rho$ and $\eta$ are described below. 

%
%

\subsection{CP Violation for Neutral Kaons}

The violation of the CP symmetry has been observed, to date, only in the 
neutral kaon system, whose mass eigenstates can be written as:
\begin{displaymath}
|K_S\rangle = p |K^0\rangle + q |\bar{K^0}\rangle 
\,\,\,\,\,\,\,\,\,\,
|K_L\rangle = p |K^0\rangle - q |\bar{K^0}\rangle.
\end{displaymath}
The relation $p\neq q$ implies the violation of CP that, in the 
Wu-Yang phase convention~\cite{wuyang}, is described by the parameter $\epsilon_K$
defined as:
\begin{displaymath}
{p \over q} = {1 + \epsilon_K \over 1 - \epsilon_K  }.
\end{displaymath}
Information from the precise measurements of the $K_S \rightarrow \pi^+ \pi^-$ and 
$K_L \rightarrow \pi^+ \pi^-$ decay rates lead to the determination~\cite{pdg}:
\begin{displaymath}
|\epsilon_K| = (2.280 \pm 0.019) \times 10^{-3}.
\end{displaymath}
The relation of $|\epsilon_K|$ with the CKM matrix parameters can be written~\cite{buras1,buras2} as:
\begin{displaymath}
|\epsilon_K| = {G_F^2 f^2_K m_K m_W^2 \over 6 \sqrt{2} \pi^2 \Delta m_K}
 B_K \left( A^2 \lambda^6 \eta \right)\times
\end{displaymath}
\begin{equation}
\times\big[ y_c \left( \eta_{ct}f_3(y_c,y_t)  \eta_{cc}\right) 
 + \eta_{tt} y_t f_2(y_t) A^2 \lambda^4 \left(1-\rho\right)\big]                           
\label{equation:ek}
\end{equation}
The functions $f_3$ and $f_2$ of the variables
$y_i = m_i^2 /m_W^2$ can be found in~\cite{ali1}.

From the value of the mass of the top quark reported by the CDF and D0 
collaborations~\cite{pdg}, $173.8 \pm 5.2\,\GeV$, and the scaling
proposed in~\cite{buras3} one obtains:

\begin{displaymath}
\overline{m_t}(m_t) = 166.8 \pm 5.3\,\GeV,
\end{displaymath}
while the mass of the charm quark is~\cite{pdg}:
\begin{displaymath}
\overline{m_c}(m_c) = 1.25 \pm 0.15\,\GeV.
\end{displaymath}
The  QCD corrections 
are calculated~\cite{buras3,buras4,buchalla1etal} to be:
\begin{displaymath}
\eta_{cc} = 1.38 \pm 0.53,\,\,\,\,
\eta_{tt} = 0.574 \pm 0.004\,\,\,\,{\mathrm{and}}\,\,\,\,
\eta_{ct} = 0.47 \pm 0.04.
\end{displaymath}
The larger theoretical uncertainty that affects this constraint is that on the ``bag'' 
parameter $B_K$, that reflects non-perturbative QCD contributions to the process. 
Using the value of the JLQCD collaboration~\cite{JLQCD},
$B_K(2\,\GeV) = 0.628\pm 0.042$, with a calculation similar to that
reported in~\cite{sharpe1}, it is possible to derive:
\begin{displaymath}
B_K  = 0.87\pm 0.14.
\end{displaymath}

The other physical constants of this and of the following constraints 
are reported in the left half of Table~1, whose numerical values not
described in the text are all from~\cite{pdg}. This constraint
has the shape of an hyperbola in the $\rho-\eta$ plane.

%
%

\subsection{Oscillations of $\boldmath{B^0_d}$ Mesons}

Neutral mesons containing a $b$ quark show a behaviour similar to neutral kaons. 
The mass difference $\Delta m_d$ of the two interaction eigenstates 
is the key feature of the physics  while the lifetime difference dominates the 
effects in the neutral kaon system.
The LEP experiments have measured $\Delta m_d$
by investigating the oscillations
of one CP eigenstate into the other~\cite{bately}:
\begin{displaymath}
\Delta m_d            = 0.466 \pm 0.019\,\mathrm{ps}^{-1}.
\end{displaymath}
The relation of $\Delta m_d$ with the CKM parameters, making
use of the Standard Model description of the box diagrams that
give rise to the mixing, is:
\begin{equation}
\Delta m_d = {G_F^2 \over 6 \pi ^2}m_W^2 m_B^2 \left(f_{B_d}\sqrt{B_{B_d}}\right)^2
\eta_B y_t f_2(y_t) A \lambda^6 \left[ \left(1-\rho\right)^2 + \eta^2\right].
\label{equation:dmd}
\end{equation}
The calculated value of the QCD correction 
is~\cite{buras3,buras4,buchalla1etal}:
\begin{displaymath}
     \eta_B  = 0.55 \pm 0.01,
\end{displaymath}
while the non-perturbative QCD parameter  $f_{B_d}\sqrt{B_{B_d}}$,
is taken as~\cite{flynn2}:
\begin{displaymath}
     f_{B_d}\sqrt{B_{B_d}} = (0.201 \pm 0.042)\,\GeV.
\end{displaymath}
This measurement of $\Delta m_d$ constraints the vertex of the unitarity triangle to a
circle in the $\rho-\eta$ plane, centred in $(1,0)$.

\subsection{Oscillations of $\boldmath{B^0_s}$ Mesons}

$B^0_s$ mesons are believed to undergo a mixing analogous to the 
$B^0_d$ ones, but their larger mass difference 
$\Delta m_s$ is responsible for faster and thus still undetected oscillations.
The combined LEP  limit is~\cite{bately}:
\begin{displaymath}
  \Delta m_s > 10.2\,\mathrm{ps}^{-1}\,\,\, (95\%\,\mathrm{C.L.}).
\end{displaymath}
The expression for $\Delta m_s$ in the Standard Model is similar to that
for  $\Delta m_d$ and their ratio yields:
\begin{equation}
\Delta m_s = \Delta m_d {1 \over \lambda^2}{m_{B_s} \over m_{B_d}} \xi^2 
{1 \over \left( 1- \rho\right)^2 + \eta^2}.
\label{equation:dms}
\end{equation}
All the theoretical uncertainties are included in the  quantity
$\xi$, evaluated as~\cite{flynn2}:
\begin{displaymath}
\xi = { f_{B_d}\sqrt{B_{B_d}} \over f_{B_s}\sqrt{B_{B_s}} } = 1.14 \pm 0.08.
\end{displaymath}

This experimental lower limit excludes all the  values of the vertex
of the unitarity triangle outside  a circle in the $\rho-\eta$ plane  with centre in $(1,0)$.


\subsection{Charmless Semileptonic b Decays}

Each of the three constraints described above is affected by a large
non-perturbative QCD uncertainty on some parameters, respectively 
 $B_K$, $f_{B_d}\sqrt{B_{B_d}}$ and $\xi$. 
It follows from the CKM matrix parametrisation~(\ref{equation:ckm}) that:
\begin{equation}
 |\mathrm{V_{ub}}|/|\mathrm{V_{cb}}| = \lambda \sqrt{\rho^2 + \eta^2}.
\label{equation:vubvcb}
\end{equation}
An experimental determination of either  $|\mathrm{V_{ub}}|$ or the ratio 
$|\mathrm{V_{ub}}|/|\mathrm{V_{cb}}|$ is  thus a  constraint unaffected by these
uncertainties.

The CLEO collaboration measured both the $|\mathrm{V_{ub}}|/|\mathrm{V_{cb}}|$
ratio and the value of $|\mathrm{V_{ub}}|$ by means, respectively, 
of the endpoint of inclusive~\cite{cleoinc2} and 
exclusive~\cite{cleoexc} charmless semileptonic b decays, obtaining the
results:
\begin{displaymath}
|\mathrm{V_{ub}}|/|\mathrm{V_{cb}}| = 0.08 \pm 0.02 \,\,\,\mathrm{and}\,\,\,
|\mathrm{V_{ub}}| = (3.3 \pm 0.2  \,^{+0.3}_{-0.4} \pm 0.7) \times 10^{-3},
\end{displaymath}
where the uncertainties on the second measurement are respectively statistical,
systematic and theoretical.
The ALEPH and L3 collaborations  have recently 
measured at LEP the  inclusive charmless semileptonic branching
fraction of beauty hadrons, $\mathrm{Br}(b\rightarrow X_u\ell\nu)$, from
which the value of $|\mathrm{V_{ub}}|$ can be extracted as in~\cite{uraltsev}.
From the experimental results:
\vspace{1ex}
\begin{center}
\begin{tabular}{rl}
ALEPH~\cite{alephvub}: & Br$(b\rightarrow X_u\ell\nu) = (1.73 \pm 0.55 \pm 0.55) \times 10^{-3}$\\
L3~\cite{l3vub}:    & Br$(b\rightarrow X_u\ell\nu) = (3.3 \pm 1.0 \pm 1.7) \times 10^{-3}$,\\
\end{tabular}
\end{center}
\vspace{1ex}
where the first uncertainty is statistical and the second systematic, the following 
average can be obtained:
\begin{displaymath}
 \mathrm{Br}(b\rightarrow X_u\ell\nu) = (1.85 \pm 0.52 \pm 0.59 )\times 10^{-3},
\end{displaymath}
with the same meaning of the uncertainties. This value makes it possible to determine
$|\mathrm{V_{ub}}|$  at LEP by means of the formula described in~\cite{uraltsev} as:
\begin{displaymath}
|\mathrm{V_{ub}}| = (4.5  \,^{+0.6}_{-0.7}\, ^{+0.7}_{-0.8} \pm 0.2)\times 10^{-3}.
\end{displaymath}
The first uncertainty is statistical, the second systematic and the third
theoretical.
The combination of this value with the CLEO exclusive one gives:
\begin{displaymath}
|\mathrm{V_{ub}}| = ( 3.8 \pm 0.6) \times 10^{-3},
\end{displaymath}
that, with the quoted value of $|\mathrm{V_{cb}}|$, yields:
\begin{displaymath}
|\mathrm{V_{ub}}|/|\mathrm{V_{cb}}| = 0.093 \pm 0.016.
\end{displaymath}

A circle in the $\rho-\eta$ plane with centre in (0,0) represents this constraint.
Figure~2a shows all the described constraints.

%
%
\section{Fit Procedure and Results}

The $\rho$ and $\eta$ parameters can be determined with the following fit procedure.
The experimental and theoretical quantities that appear in the formulae 
describing the constraints have
been fixed to their central values if their errors were below the 2\%, and are
reported in the left half of~Table~1. The quantities affected by larger
errors have been used as additional parameters of the fit, including a constraint
on their value. 
The following expression has then been minimised using the MINUIT package~\cite{minuit}:
\begin{displaymath}
\chi^2 =  {\left(\widehat{A} - A\right)^2 \over \sigma_{A}^2} +      
          {\left(\widehat{m_c} - m_c\right)^2 \over \sigma_{m_c}^2} +
          {\left(\widehat{m_t} - m_t\right)^2 \over \sigma_{m_t}^2} + 
          {\left(\widehat{B_K} - B_K\right)^2 \over \sigma_{B_K}^2} +
          {\left(\widehat{\eta_{cc}} -\eta_{cc} \right)^2 \over \sigma_{\eta_{cc}}^2} +           
\end{displaymath}
\begin{displaymath}
          +{\left(\widehat{\eta_{ct}} -\eta_{ct} \right)^2 \over \sigma_{\eta_{ct}}^2} + 
           {\left(\widehat{f_{B_d}\sqrt{B_{B_d}}} - f_{B_d}\sqrt{B_{B_d}}\right)^2 
                 \over \sigma_{f_{B_d}\sqrt{B_{B_d}}}^2} +
           {\left(\widehat{\xi} - \xi\right)^2 \over \sigma_{\xi}^2} + 
           {\left(\widehat{|\mathrm{V_{ub}}|\over|\mathrm{V_{cb}}|} 
               - {|\mathrm{V_{ub}}|\over|\mathrm{V_{cb}}|}\right)^2 \over \
             \sigma_{{|\mathrm{V_{ub}}| \over |\mathrm{V_{cb}}|}}^2} +
\end{displaymath}
\begin{displaymath}
          +{\left(\widehat{|\epsilon_K|} - |\epsilon_K| \right)^2  
          \over  \sigma_{|\epsilon_K|}^2} +
          {\left(\widehat{\Delta m_d} - \Delta m_d\right)^2
          \over  \sigma_{\Delta m_d}^2}+
        {\left(1-{\cal{A}}\left(\Delta m_s\right)\right)^2
          \over \sigma_{{\cal{A}}\left(\Delta m_s \right)}^2}.
\end{displaymath}
The symbols with a hat represent the reference values measured or 
calculated for a given physical quantity, as listed in the right half
of Table~1, while the corresponding $\sigma$ are their errors. 
The parameters of the fit are $\rho$, $\eta$,  $A$, $m_c$, $m_t$, $B_K$, 
$\eta_{cc}$, $\eta_{ct}$, $f_{B_d}\sqrt{B_{B_d}}$ and 
$\xi$, that are used to calculate the values of $|\epsilon_K|$,  $\Delta m_d$,  
$\Delta m_s$ and $|\mathrm{V_{ub}}|/\mathrm{V_{cb}}|$
by means of the formulae~(\ref{equation:ek}),~(\ref{equation:dmd}),~(\ref{equation:dms}) 
and~(\ref{equation:vubvcb}).

The $\Delta m_s$ limit has been included in the $\chi^2$ as suggested in~\cite{paganini}.
The results of the search for $B^0_s$ oscillations are combined~\cite{bately} 
in terms of the oscillation amplitude ${\cal{A}}$~\cite{moser},
a parameter that is zero if no oscillations are observed and is compatible with one 
in presence of a signal.
The information on the dependence of both ${\cal{A}}$ and its error on $\Delta m_s$ is fully taken into
account by comparing ${\cal{A}}$
with one within its error.

\begin{table}[t] 
  \begin{center}
   \caption{Values of the physical constants (left) and parameters of the fit (right).}
    \begin{tabular}{|rcl|rcl|}
     \hline
     $\lambda               $& = & $0.2196(23)$ &                           $A                     $& = & $0.819(35)$   \\    
     $G_F                   $& = & $1.16639(1)\times 10^{-5}$\,\GeV$^{-2}$& $\eta_{ct}             $& = & $0.47(4)$   \\ 
     $f_K                   $& = & $0.1598(15)$\,\GeV &                     $\eta_{cc}             $& = & $1.38(53)$  \\  
     $\Delta m_K            $& = & $0.5304(14)\times 10^{-2}$\,ps$^{-1}$ &  $\overline{m_c}(m_c)        $& = & $1.25(15)$\,\GeV   \\
     $m_K                   $& = & $0.497672(31)$\,\GeV &                   $\overline{m_t}(m_t)        $& = & $166.8(5.3)$\,\GeV  \\ 
     $m_W                   $& = & $80.375(64)$\,\GeV &                     $f_{B_d}\sqrt{B_{B_d}} $& = & $0.201(42)$\,\GeV  \\ 
     $m_{B_d}               $& = & $5.2792(18)$\,\GeV &                     $B_K                   $& = & $0.87(14)$ \\  
     $m_{B_s}               $& = & $5.3962(20)$\,\GeV &                     $\xi                   $& = & $1.14(8)$ \\  
     $m_B                   $& = & $5.290(2)$\,\GeV &                       $|\epsilon_K|          $& = & $2.280(19)\times 10^{-3}$  \\ 
     $\eta_B                $& = & $0.55(1)$ &                              $\Delta m_d            $& = & $0.466(19)$\,ps$^{-1}$ \\    
     $\eta_{tt}             $& = & $0.574(4) $&                             $|\mathrm{V_{ub}}|/|\mathrm{V_{cb}}| $& = & $0.093(16) $ \\  
     \hline    
   \end{tabular}
 \end{center}
\end{table}

The results of the fit are the following:
\begin{displaymath}
    \rho =0.155 _{-0.105} ^{+0.115}\,\,\,\,\,
    \eta =0.383 _{-0.060} ^{+0.063}.
\end{displaymath}
The 95\% Confidence Level regions for $\rho$ and $\eta$ are:
\begin{displaymath}
    -0.10 < \rho < 0.35 \,\,\,\,\,
     0.27 < \eta < 0.50 \,\,\,(\mathrm{95\% C.L.}).
\end{displaymath}
Figure~2b shows these confidence regions  together
with the favoured unitarity triangle, also superimposed on to the constraints
of Figure~2a.

From these results it is also possible to determine  the value of the angles
of the unitarity triangle as:
\begin{displaymath}
    \sin{2\alpha}  =0.08 _{-0.50}  ^{+0.43}\,\,\,\,\,
    \sin{2\beta}   =0.75 \pm 0.10\,\,\,\,\,
    \gamma         =68 \pm 15^\circ,
\end{displaymath}
and, at the 95\% of Confidence Level:
\begin{displaymath}
   -0.75 <  \sin{2\alpha} < 0.94 \,\,\,\,\,
   0.54 <  \sin{2\beta}  < 0.91 \,\,\,\,\,
   43^\circ  <   \gamma       < 107^\circ  \,\,\,\,\,(\mathrm{95\% C.L.}).  
\end{displaymath}
The accuracy on $\sin{2\beta}$ from these indirect studies is already 
of the same level of that expected to be achieved with the direct measurement 
at B-factories due to start in the near future.

%
%
\section{Consequences of the Fit}

As different models have been proposed to explain the CP violation
in the neutral kaon system, it
is of interest to remove from the fit the constraints
related to this process and then investigate the compatibility of
$\eta$ with zero. This procedure yields the following results, also displayed in Figure~2c:
\begin{displaymath}
    \rho =0.126 _{-0.084} ^{+0.141}\,\,\,\,\,
    \eta =0.404 _{-0.95} ^{+0.70};
\end{displaymath}
$\eta$ is not compatible with zero at the 95\% of
Confidence Level either:
\begin{displaymath}
    -0.120 < \rho < 0.366 \,\,\,\,\,0.176 < \eta < 0.540 \,\,\,\,(\mathrm{95\% C.L.}).
\end{displaymath}

It is interesting to remove from the fit the constraints on the parameters affected 
by the largest theory uncertainty:  $B_K$ and $f_{B_d}\sqrt{B_{B_d}}$.
This allows to extract information on their values and to determine
$\rho$ and $\eta$ independently of these uncertainties.
This method for the parameter $B_K$ yields:
\begin{displaymath}
    \rho =0.126 _{-0.084} ^{+0.140}\,\,\,\,\,
    \eta = 0.404 _{-0.095} ^{+0.069}\,\,\,\,\,
    B_K  = 0.76 _{-0.16} ^{+0.33}.
\end{displaymath}
The favoured central value of $B_K$ is
lower than the input one, as suggested by other analyses~\cite{buras4}.
The same procedure with  $f_{B_d}\sqrt{B_{B_d}}$ as a free parameter leads to 
the results:
\begin{displaymath}
    \rho =0.191 _{-0.134} ^{+0.121}\,\,\,\,\,
    \eta = 0.378 _{-0.060} ^{+0.065}\,\,\,\,\,
    f_{B_d}\sqrt{B_{B_d}} = 0.222 _{-0.026} ^{+0.030}\,\GeV.
\end{displaymath}
The value of $f_{B_d}\sqrt{B_{B_d}}$ agrees
with the predicted one and shows a smaller uncertainty.

The $\Delta m_s$ constraint has a big impact on the $\rho$ negative
error as can be observed by removing it from the fit, what gives:
\begin{displaymath}
    \rho =0.155 _{-0.204} ^{+0.115}\,\,\,\,\,
    \eta =0.382 _{-0.060} ^{+0.064}.
\end{displaymath}
\begin{displaymath}
    -0.46 < \rho < 0.35 \,\,\,\,\,0.27 < \eta < 0.51 \,\,\,\,(\mathrm{95\% C.L.}),
\end{displaymath}
as shown in Figure~2d. The confidence regions for $\Delta m_s$ can be extracted as:
\begin{displaymath}
    \Delta m_s = 15.0 _{-4.4} ^{+4.7}\,\mathrm{ps}^{-1} 
\end{displaymath}
\begin{displaymath}
     5.6\,\mathrm{ps}^{-1} < \Delta m_s   < 23.9\,\mathrm{ps}^{-1} \,\,\,(\mathrm{95\% C.L.}),
\end{displaymath}
to be compared with the LEP expected sensitivity: $\Delta m_s < 13\,\mathrm{ps}^{-1}$~\cite{bately}.

%
%
\section{Summary}

The precise LEP measurements of $\Delta m_d$, the  limits
on $\Delta m_s$ and the recent determination of  $|\mathrm{V_{ub}}|$
 improve the constraints on the CKM matrix elements.

With a  fit to these constraints it is possible to determine
the vertex of the unitarity triangle:
\begin{displaymath}
    \rho =0.155 _{-0.105} ^{+0.115}\,\,\,\,\,
    \eta =0.383 _{-0.060} ^{+0.063},
\end{displaymath}
what yields for its angles:
\begin{displaymath}
    \sin{2\alpha}  =0.08 _{-0.50}  ^{+0.43}\,\,\, 
    \sin{2\beta}   =0.75 \pm 0.10\,\,\,
    \gamma         =68 \pm 15^\circ. 
\end{displaymath}
The precision on $\sin{2\beta}$ is comparable with the 
expected one from direct measurement at  future B factories.
Moreover, the presence of a complex phase in the matrix is established
at more than the  95\% of Confidence Level even removing from the
analysis the constraints from the CP violation in the neutral kaon system.
Assuming the Standard Model, the fit also gives:
\begin{displaymath}
    f_{B_d}\sqrt{B_{B_d}} = 0.222 _{-0.026} ^{+0.030}\,\GeV\,\,\,\,\,
    \Delta m_s = 15.0 _{-4.4} ^{+4.7}\,\mathrm{ps}^{-1}. 
\end{displaymath}

%
%

\section*{Acknowledgements}

I am grateful to the organisers of this workshop for having invited me to
present these results and to Joachim Mnich for the useful discussions on the fit
procedure.

%
%

\begin{figure}[p]
  \begin{center}
    \begin{tabular}{cc}
      \mbox{\epsfysize=6cm\epsffile{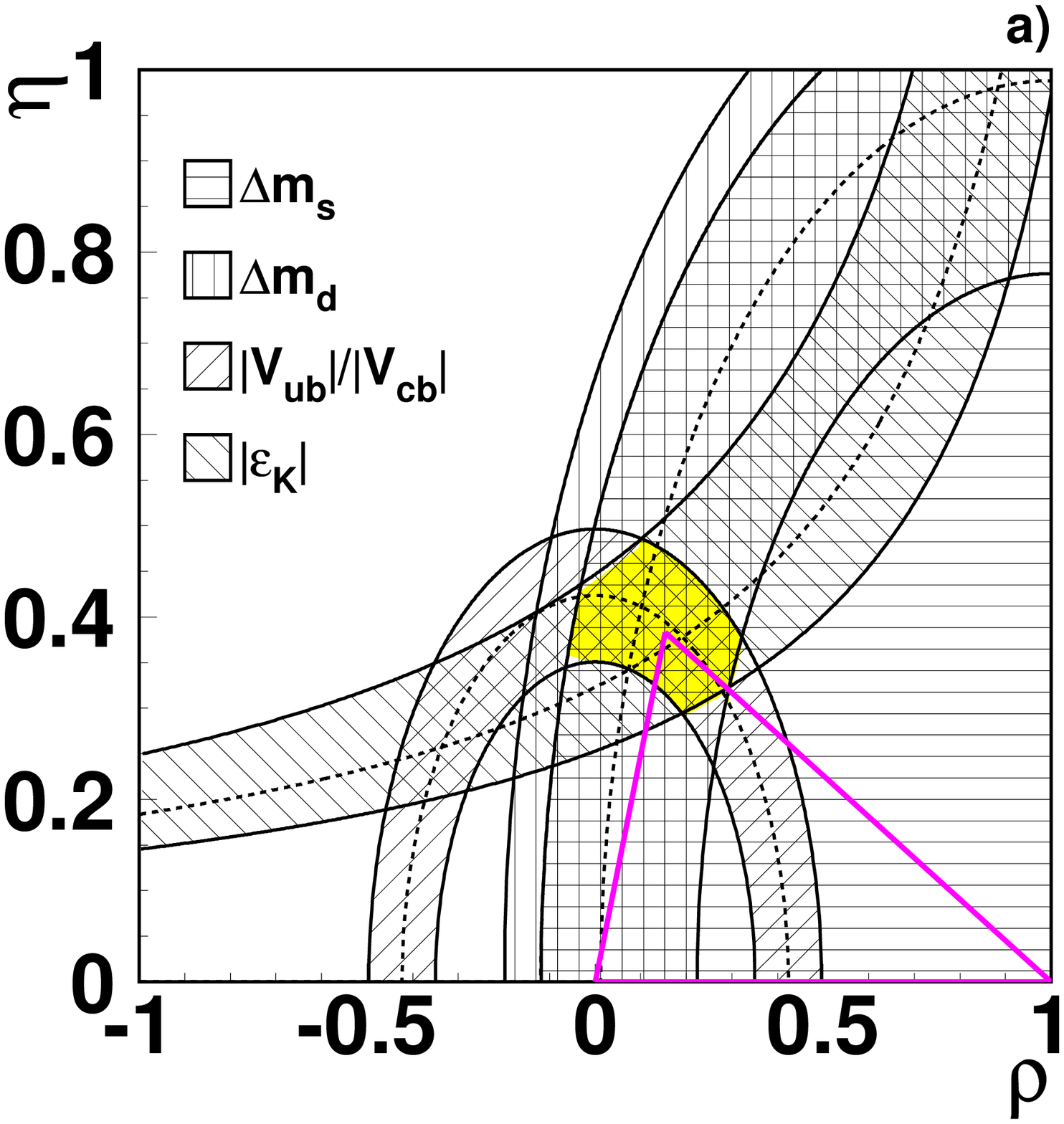}} &
      \mbox{\epsfysize=6cm\epsffile{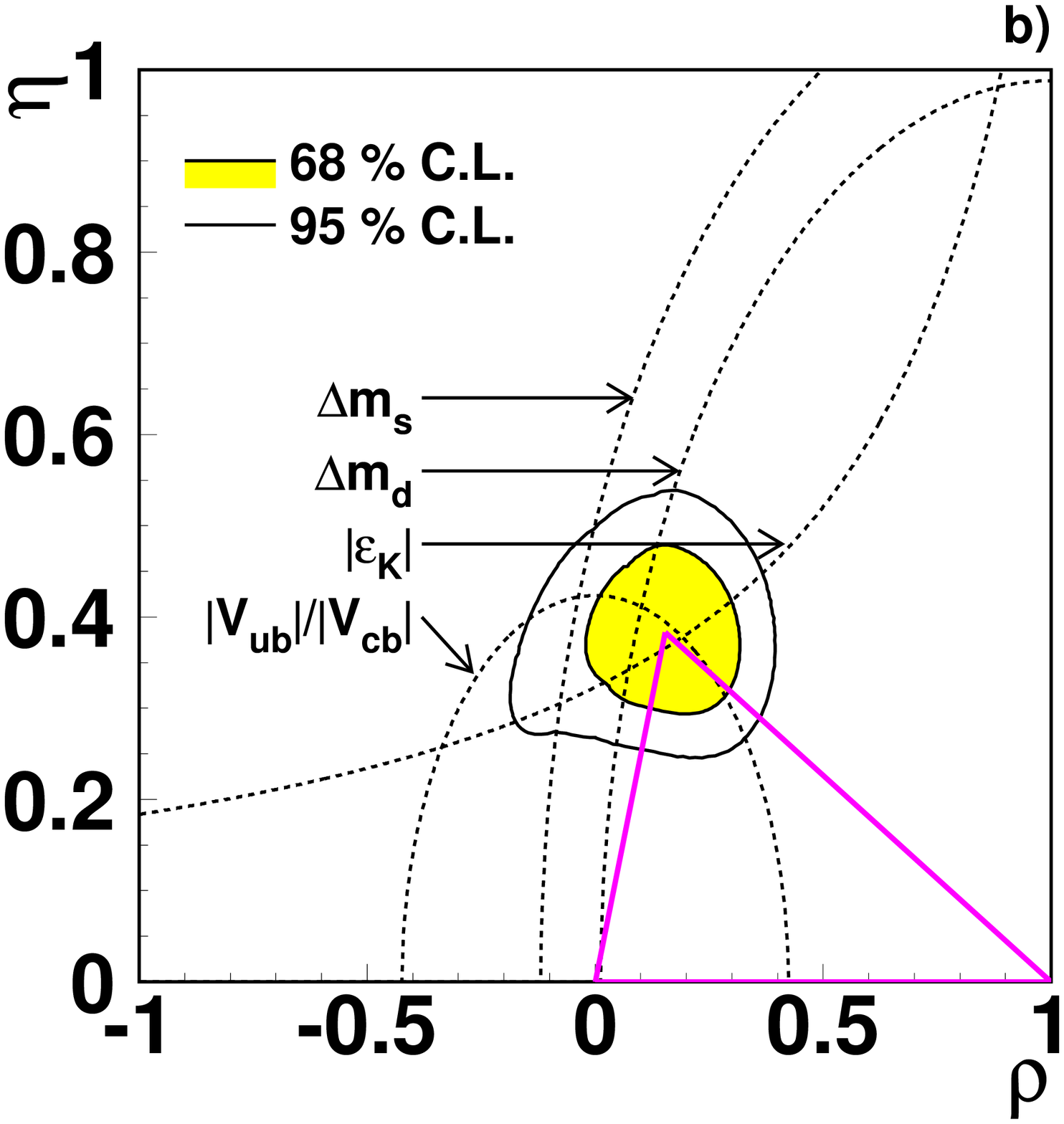}} \\
      \mbox{\epsfysize=6cm\epsffile{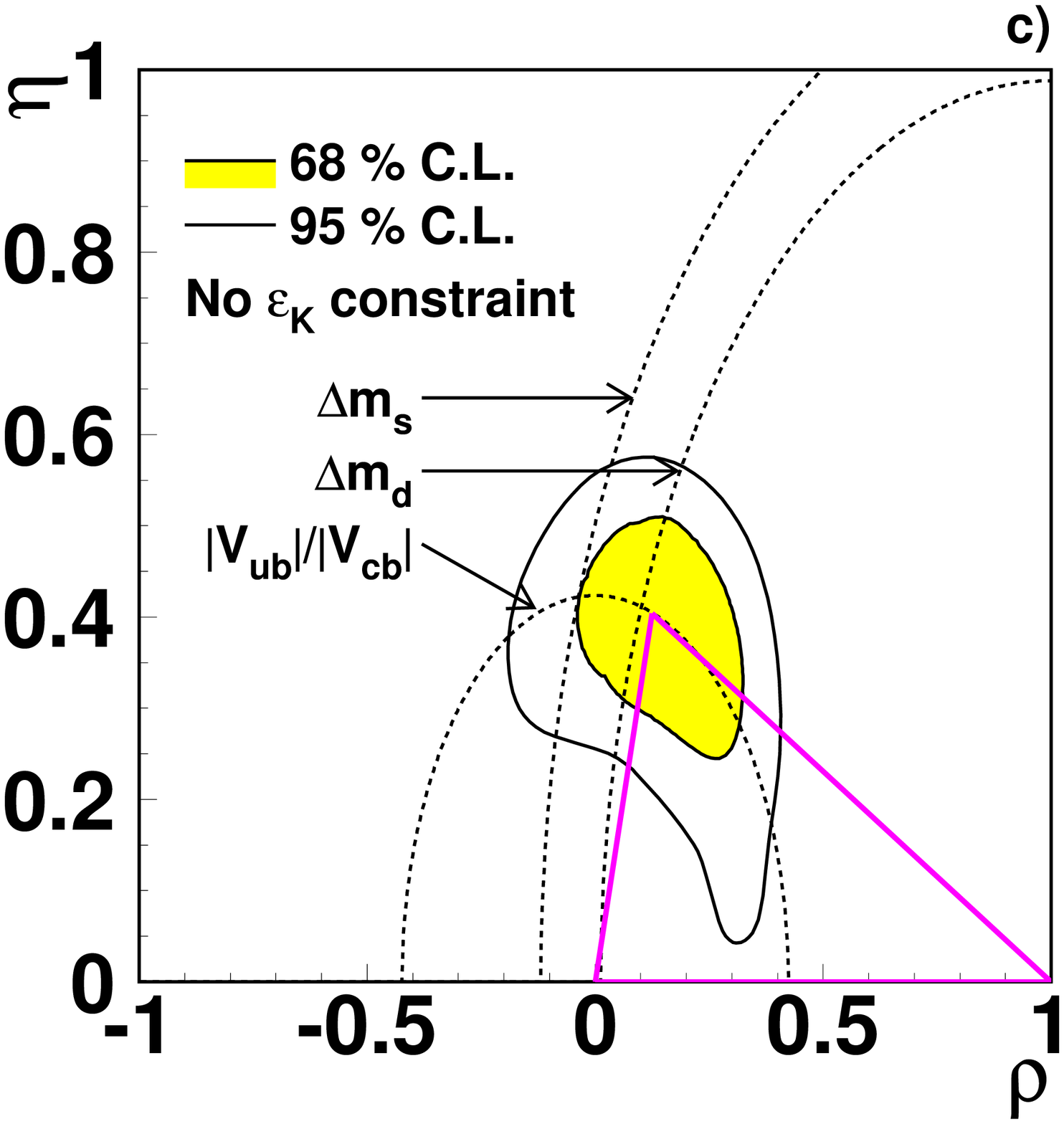}} &
      \mbox{\epsfysize=6cm\epsffile{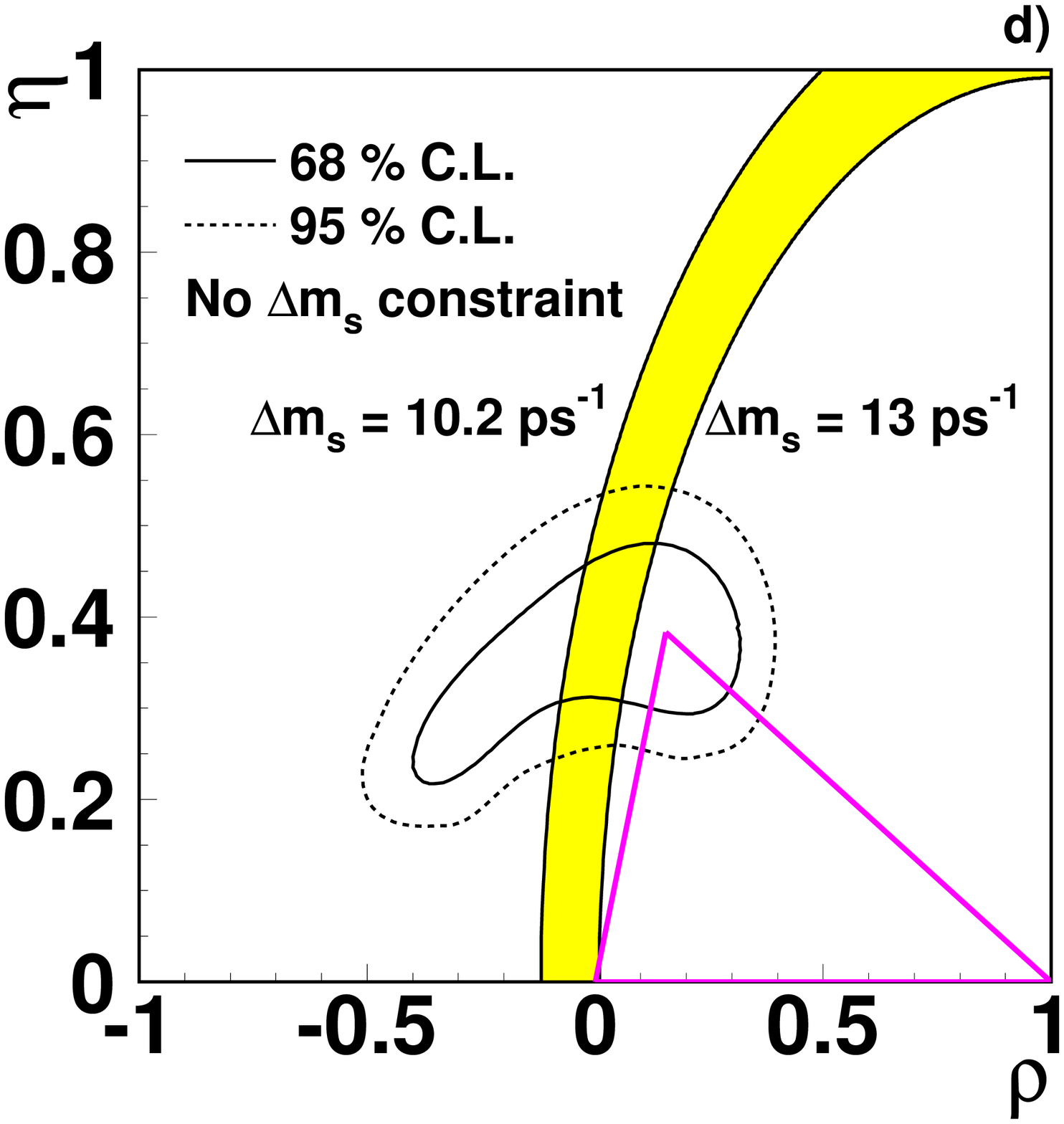}} \\
    \end{tabular}
    \caption{
      a) The constraints and the favoured unitarity triangle; the constraint
      coming from $B^0_s$ oscillations is a limit at 95\% of Confidence
      Level, while the others represent a $\pm 1 \sigma$ variation of the experimental
      and theoretical parameters entering the formulae in the text.
      b) c) and d)
      The fit unitarity triangle and the confidence regions for its vertex in the
      following assumptions:
      b) the fit to all the data,
      c) the constraints from the neutral kaon system are not applied,
      d) the constraint from the $B^0_s$ oscillations is not applied. The band
      in d) displays the values of $\rho$ and $\eta$ corresponding to a value of $\Delta m_s$ 
      between the  lower limit and the LEP sensitivity.
      The central values of the constraints and the $\Delta m_s$ limit are also shown 
      in b) and c).}
  \end{center}
\end{figure}

%
%

\end{document}